\documentclass[aps,  prb,twocolumn,
  superscriptaddress,  floatfix,  10pt         
]{revtex4-2}
\usepackage{float}
\usepackage{graphicx}
\usepackage{amsmath,amssymb}
\usepackage{tikz}
\usepackage{braket}
\usepackage{pgfplots}
\pgfplotsset{compat=1.17}
\usepackage[version=4]{mhchem}
\usepackage{physics}
\usepackage{subfigure}
\usepackage{enumitem}
\usepackage{mathrsfs}
\usepackage{mathtools}
\usepackage{subcaption}

\definecolor{phase1a}{RGB}{29,145,192}   % dark steel blue
\definecolor{phase1b}{RGB}{184,207,229}  % pale steel blue
\definecolor{phase2}{RGB}{236,203,204}   % light peach/pink (YSR)
\definecolor{phase3b}{RGB}{115,188,79}   % forest green
\definecolor{DarkGreen}{rgb}{0.0,0.5,0.0}
\definecolor{BrickRed}{rgb}{0.2,0.2,0.8}

\usetikzlibrary{
  shapes.geometric,
  shadows,
  patterns,
  perspective,
  decorations,
  decorations.text
  }

\usetikzlibrary{decorations.markings,calc,decorations.pathreplacing}

\tikzset{
  baseline/.style = {line width=0.8pt, draw=black!70, line cap=round},
  tick/.style     = {line width=0.9pt, draw=black, line cap=round},
  site/.style     = {fill=black, draw=none},
  Jbond/.style    = {line width=1.4pt, draw=green!70!black, dotted, line cap=round},
  gbond/.style    = {line width=1.4pt, draw=cyan!80!black, line cap=round},
  Jlab/.style     = {font=\bfseries\footnotesize, text=green!60!black},
  glab/.style     = {font=\bfseries\footnotesize, text=cyan!80!black},
  upA/.style      = {-stealth, line width=0.5mm, draw=red!80!black},
  dnA/.style      = {-stealth, line width=0.5mm, draw=blue!80!black}
}

\newcommand{\nicechain}[4]{%
  \begin{scope}[shift={#1}]
    \def\dx{1.20}\def\Larr{1.00}\def\yoff{0.50}\def\rsite{0.12}\def\ticklen{0.17}
    \newcount\Np \Np=0
    \foreach \s [count=\j] in {#2} {\global\Np=\j}
    \newcount\N \N=\numexpr\Np-1\relax
    \ifnum\Np>1
      \draw[Jbond] ({1*\dx},0) -- ({2*\dx},0);
      \path ($({1*\dx},0)!0.5!({2*\dx},0)$) ++(0,0.20) node[Jlab]{J};
    \fi
    \ifnum\N>2
      \foreach \j in {2,...,\numexpr\N-1\relax}{
        \pgfmathtruncatemacro{\jp}{\j+1}
        \draw[gbond] ({\j*\dx},0) -- ({\jp*\dx},0);
        \path ($({\j*\dx},0)!0.5!({\jp*\dx},0)$) ++(0,0.20) node[glab]{g};
      }
    \fi
    \ifnum\Np>2
      \draw[Jbond] ({\N*\dx},0) -- ({\Np*\dx},0);
      \path ($({\N*\dx},0)!0.5!({\Np*\dx},0)$) ++(0,0.20) node[Jlab]{J};
    \fi
    \foreach \s [count=\j] in {#2}{
      \draw[tick] ({\j*\dx},-\ticklen) -- ({\j*\dx},\ticklen);
      \fill[site] ({\j*\dx},0) circle (\rsite);
      \ifnum\s=1
        \draw[upA] ({\j*\dx},-\yoff) -- ++(0,\Larr);
      \else
        \draw[dnA] ({\j*\dx},\yoff) -- ++(0,-\Larr);
      \fi
    }
  \end{scope}
}
\usepackage{xcolor}

\definecolor{linkcolor}{RGB}{0,0,255}      % blue
\definecolor{citecolor}{RGB}{0,128,0}     % green
\definecolor{urlcolor}{RGB}{255,0,0}      % red

\usepackage{hyperref}
\hypersetup{
  colorlinks   = true,
  linkcolor    = linkcolor,
  citecolor    = citecolor,
  urlcolor     = urlcolor,
  linktoc      = all,
  pdfborder    = {0 0 0},
}

\definecolor{dy}{rgb}{0.9,0.9,0.4}
\definecolor{dr}{rgb}{0.95,0.65,0.55}
\definecolor{db}{rgb}{0.5,0.8,0.9}
\definecolor{dg}{rgb}{0.2,0.9,0.6}
\definecolor{Navy}{rgb}{0.2,0.2,0.6}
\definecolor{DarkGreen}{rgb}{0.1,0.4,0.1}
\definecolor{phaseK}{HTML}{CC79A7}
\definecolor{phaseY}{HTML}{56B4E9}
\definecolor{phaseU}{HTML}{E69F00}
\definecolor{phaseF}{HTML}{23EB91}

\allowdisplaybreaks[4]
\begin{document}

\title{Breakdown of Monotonic Impurity Entropy Flow in $\mathscr{PT}$-Symmetric Multichannel Kondo Systems}

\author{Pradip Kattel}
\email{pradip.kattel@unige.ch}
\affiliation{Department of Quantum Matter Physics, University of Geneva, Quai Ernest-Ansermet 24, 1211 Geneva, Switzerland}

\author{Abay Zhakenov}
\author{Natan Andrei}
\affiliation{Department of Physics and Astronomy, Center for Materials Theory, Rutgers University, Piscataway, New Jersey 08854, USA}

\begin{abstract}
We study a $\mathscr{PT}$-symmetric  non-Hermitian multichannel Kondo model consisting of a pair of spin-$\frac12$ impurities coupled to $n$ conduction-electron channels through complex-conjugate Kondo couplings. The impurity renormalization-group (RG) flow is characterized by two invariants: the Kondo scale $T_K$, generalizing the conventional Kondo temperature, and a dimensionless parameter $\alpha$ measuring the departure from Hermiticity. As $\alpha$ increases, the exact Bethe Ansatz solution reveals four impurity phases: overscreened Kondo $(0<\alpha<\pi/2)$, zero mode $(\pi/2<\alpha<n\pi/2)$, Yu--Shiba--Rusinov (YSR) $(n\pi/2<\alpha<(\frac{n}{2}+1)\pi)$, and local moment $(\alpha>(\frac{n}{2}+1)\pi)$. The Kondo, zero-mode, and local-moment phases are $\mathscr{PT}$-unbroken, whereas the YSR phase spontaneously breaks $\mathscr{PT}$ symmetry. Using a generalized thermodynamic Bethe Ansatz, we determine the impurity free energy and Affleck--Ludwig $g$-function throughout the $\mathscr{PT}$-unbroken phases. In the Kondo phase, the defect RG flow connects the ultraviolet and infrared conformal fixed points, with the impurity entropy flowing from $2\ln2$ to $2\ln\left[2\cos\left(\frac{\pi}{n+2}\right)\right]$, in agreement with defect conformal field theory. In the zero-mode phase, zero-energy fundamental and higher-order impurity strings reorganize the spectrum into two and three excitation towers; in the YSR phase, spontaneous $\mathscr{PT}$ symmetry breaking produces a complex spectrum beyond the scope of our thermodynamic Bethe Ansatz; and in the local-moment phase, the RG flow becomes cyclic, returning to the unscreened local-moment fixed point. We conjecture that RG irreversibility, and hence a generalized Affleck--Ludwig $g$-theorem, survives throughout the Kondo phase $0<\alpha<\pi/2$, where excitations remain organized into a single tower. Our exact solution shows, however, that neither a real spectrum nor ultraviolet and infrared defect entropies consistent with defect CFT are sufficient to guarantee RG irreversibility: in the zero-mode phase, the impurity entropy develops intermediate overshoots and undershoots between distinct ultraviolet and infrared fixed-point values, whereas in the local-moment phase it returns to the UV value $2\ln2$ through intermediate overshoots and undershoots.
\end{abstract}

\maketitle

Quantum impurity systems provide canonical realizations of integrable defect renormalization-group (RG) flows. In the multichannel Kondo effect, a localized spin is screened by conduction electrons through a defect RG flow connecting ultraviolet and infrared conformal defect fixed points, giving rise to universal non-Fermi-liquid behavior~\cite{wilson1975renormalization,andrei1983solution,tsvelick1983exact,affleck1995conformal,gaiotto2021integrable}. In the chiral formulation, the impurity is described by an integrable Kondo line defect in an $SU(2)_n$ Wess--Zumino--Witten (WZW) conformal field theory (CFT), whose RG flow interpolates between conformal defect fixed points~\cite{gaiotto2021integrable}.

Recent work has established a broad framework for non-Hermitian conformal and defect criticality, including complex conformal field theories, non-unitary conformal interfaces~\cite{furuta2026complex,tang2026exactly}, lattice realizations of topological defects, boundary criticality, and holographic $\mathscr{PT}$-symmetric defect theories~\cite{Liu2026ExtractingBC,sinha2026lattice,Maeda2026Holographic,castro2017irreversibility}.

In this Letter, we investigate a $\mathscr{PT}$-symmetric multichannel Kondo line defect in a chiral conformal field theory, generalizing the non-Hermitian Kondo model of Refs.~\cite{nakagawa2018non,kattel2025dissipation,kattel2025spin,kattel2026monotonic,burke2025non} to multiple channels~\cite{yi2026interplay,yi2026non}. Unlike previous non-Hermitian conformal defects, which remain critical~\cite{Maeda2026Holographic,sinha2026lattice,furuta2026complex}, our defect is perturbed by a classically marginal operator that becomes marginally relevant, generating an integrable defect RG flow. A microscopic realization is provided by a pair of spin-$\frac12$ impurities coupled through complex-conjugate Kondo interactions to $n$ channels of conduction electrons,
\begin{align}
H=& -i v_F \sum_{a=1}^{n}\sum_{\sigma} \int dx\, \psi^\dagger_{a\sigma}(x)\partial_x\psi_{a\sigma}(x)\nonumber\\
&\quad+\lambda\mathbf S_1\cdot\mathbf J(x_1)+\lambda^*\mathbf S_2\cdot\mathbf J(x_2),
\end{align}
where $\mathbf J(x)=\frac12\sum_a\psi^\dagger_{a\alpha}(x)\boldsymbol{\sigma}_{\alpha\beta}\psi_{a\beta}(x)$ is the $SU(2)_n$ current, and the impurities $\mathbf S_{1,2}$ are located at $x_{1,2}$ on a ring of circumference $L$. The complex-conjugate couplings render the Hamiltonian $\mathscr{PT}$ symmetric. Since our focus is the chiral defect CFT, the Hamiltonian should be regarded as an integrable realization of the defect theory. Consistent with the chiral formulation~\cite{affleck1995conformal,gaiotto2021integrable}, it retains only forward scattering, with impurity backscattering and direct impurity--impurity interactions absent. As a result, the energies are independent of the impurity positions~\cite{andrei1980diagonalization}.

The exact solution follows from a generalized thermodynamic Bethe Ansatz~\cite{yang1969thermodynamics,takahashi1999thermodynamics,andrei1983solution,zhakenov2025thermodynamics,kattel2026thermodynamics}, yielding the defect free energy and Affleck--Ludwig $g$-function. The impurity free energy and the corresponding $g$-function are defined by
\begin{equation}
F_{\rm imp}(T)
\equiv F_{\rm total}(T)-F_0(T)
=-T\ln g_{\rm imp}(T),
\end{equation}
with $F_{\rm total}(T)$ the free energy of the full system including the two impurities and $F_0(T)$ that of the corresponding impurity-free system. Introducing the effective couplings $\tilde c=ce^{i\phi}=2\lambda/\left(1-\frac34\lambda^2\right)$ and $\tilde c^{*}=ce^{-i\phi}=2\lambda^{*}/\left(1-\frac34(\lambda^{*})^2\right)$~\cite{andrei1983solution,kattel2025dissipation}, we define the RG invariant $\alpha=\frac{\pi|\sin\phi|}{c}$, which measures the departure from Hermiticity and determines the impurity phase diagram~\cite{kattel2025dissipation}.

As we show below, for $0<\alpha<\frac{\pi}{2}$ the spectrum remains real, and the defect entropy flows from $2\ln2$ to $2\ln d_{\frac12}$, where $d_{\frac12}=2\cos\left(\frac{\pi}{n+2}\right)$ is the quantum dimension of the spin-$\frac12$ primary of the $SU(2)_n$ WZW model, in agreement with the conformal defect fixed point. Notably, such fractional entropy has recently been measured experimentally in Hermitian Kondo systems~\cite{child2022entropy,piquard2026experimental}. The $g$-function decreases monotonically, continuously extending the conventional multichannel Kondo effect. For $\frac{\pi}{2}<\alpha<\frac{n\pi}{2}$, the $g$-function becomes nonmonotonic despite connecting the same ultraviolet and infrared fixed points. This nonmonotonicity originates from the appearance of zero-energy impurity strings, which reorganize the excitation spectrum into multiple towers.

These results follow from the Bethe Ansatz equations (derived in the End Matter):
\begin{equation}
e^{i k_{j} L} =\prod_{\gamma=1}^{M} \frac{\Lambda_{\gamma}-1+i \frac{cn}{2}}
     {\Lambda_{\gamma}-1-i \frac{cn}{2}},
\end{equation}
where $\Lambda_\gamma$ are rapidities and $M$ denotes the number of spin flips. The rapidities satisfy

\begin{align}
&\prod_{\delta=1,\delta\neq\gamma}^{M}
\frac{\Lambda_{\delta}-\Lambda_{\gamma}+ic}
     {\Lambda_{\delta}-\Lambda_{\gamma}-ic}
=
\left(
\frac{\Lambda_{\gamma}-1-i\frac{cn}{2}}
     {\Lambda_{\gamma}-1+i\frac{cn}{2}}
\right)^{N^e}
\nonumber\\
&~\times
\left(
\frac{\Lambda_{\gamma}-1+e^{-i\phi}-i\frac{c}{2}}
     {\Lambda_{\gamma}-1+e^{-i\phi}+i\frac{c}{2}}
\right)
\left(
\frac{\Lambda_{\gamma}-1+e^{i\phi}-i\frac{c}{2}}
     {\Lambda_{\gamma}-1+e^{i\phi}+i\frac{c}{2}}
\right),
\end{align}
where $N^e$ is the number of conduction electrons. The last two factors describe scattering from the two impurities with complex-conjugate couplings
$ \tilde c $ and $ \tilde c^{*}$. For $\phi=0$, the equations reduce to those of the Hermitian multichannel Kondo problem with two impurities.

\begin{figure*}
    \centering
    \begin{tikzpicture}[scale=1.1, every node/.style={scale=1.1}]

% ===== Background phases =====
\shade[left color=phase1a, right color=phase1b] (-0.0,-0.05) rectangle (2.5,4.565);        % Kondo (blue)
\shade[left color=phase1b, right color=phase2] (2.5,-0.05) rectangle (7.5,4.565);          % Zero mode (blue→pink)
\fill[phase2] (7.5,-0.05) rectangle (12.5,4.565);                                           % YSR (solid pink)
\shade[left color=phase2, right color=phase3b] (12.5,-0.05) rectangle (15.65,4.565);        % Local moment (pink→green)

% ===== Axes and x-ticks =====
\draw [thick,black,<->] (-0.0,4.6)--(-0.0,-0.1)--(15.65,-0.1);
\node[rotate=90] at (-0.175,3) {\small $E - E_0$};
\node at (15.5,-0.3) {\small $\frac{\alpha}{\pi}$};
\node at (2.5,-0.3) {\small $\frac{1}{2}$};
\node at (-0.0,-0.3) {\small $0$};
\node at (5.0,-0.3) {\small $1$};
\node at (7.5,-0.3) {\small $\frac{n}{2}$};
\node at (10.0,-0.3) {\small $\frac{n+1}{2}$};
\node at (12.5,-0.3) {\small $\frac{n}{2}+1$};

% ===== Reference line =====
\draw [dashed,DarkGreen] (-0.0,0.4)--(15.5,0.4);
\node[rotate=90] at (-0.175,0.7) {\small $|E^{(1)}+E^{(2)}|$};
% ===== Vertical phase boundaries (full height) =====
\foreach \x in {0.65,2.5,5.0,7.5,10.0,12.5}
  \draw[dashed,gray,thick] (\x,0)--(\x,4.565);

% ===== Phase labels =====
\node at (1.19,4.35) { Kondo phase};
\node at (5.0,4.35) { Zero mode phase};
\node at (10.00,4.35) { YSR phase};
\node at (14.07,4.35) {Local moment phase};
\node at (4.4,4) {(I)};
\node at (6.2,4) {(II)};
\node at (9,4) {(I)};
\node at (11,4) {(II)};

\node at (3.25,3.4) {$\mathscr{T}_1$};
\node at (4.25,2.0) {$\mathscr{T}_2$};

\node at (5.65,2.6) {$\mathscr{T}_1$};
\node at (6.35,2.2) {$\mathscr{T}_2$};
\node at (7.1,1.65) {$\mathscr{T}_3$};

\node at (7.9,3.8) {$\mathscr{T}_1$};
\node at (8.75,2.4) {$\mathscr{T}_2$};
\node at (9.55,1.2) {$\mathscr{T}_3$};

\node at (10.45,3.2) {$\mathscr{T}_1$};
\node at (11.25,1.7) {$\mathscr{T}_2$};
\node at (11.95,1.2) {$\mathscr{T}_3$};

\node at (12.95,2.9) {$\mathscr{T}_1$};
\node at (13.98,1.95) {$\mathscr{T}_2$};
\node at (14.9,1.0) {$\mathscr{T}_3$};

% ===== Helpers for rungs =====
\newcommand{\continuum}[4]{%
  \foreach \i in {0,...,#4} {%
    \pgfmathsetmacro\y{#2 + 0.1*\i}%
    \draw[thick,black] (#1-0.25,\y)--(#1+0.25,\y);%
  }%
}
\newcommand{\bound}[2]{\draw[thick,black] (#1-0.25,#2)--(#1+0.25,#2);}

% ===== Towers with horizontal rungs =====
% Kondo
\draw[ultra thick,BrickRed] (0.85,0) -- (0.85,4.1);
\bound{0.85}{0.0}
\continuum{0.85}{0.0}{0}{41}

% Zero mode (left half)
\draw[ultra thick,BrickRed] (3.25,0) -- (3.25,3.1);
\bound{3.25}{0.0}
\continuum{3.25}{0.0}{0}{31}

\draw[ultra thick,BrickRed] (4.25,0) -- (4.25,1.7);
\bound{4.25}{0.0}
\continuum{4.25}{0.0}{0}{17}

% Zero mode (right half)
\draw[ultra thick,BrickRed] (5.65,0) -- (5.65,2.3);
\bound{5.65}{0.0}
\continuum{5.65}{0.0}{0}{23}

\draw[ultra thick,BrickRed] (6.35,0) -- (6.35,1.9);
\bound{6.35}{0.0}
\continuum{6.35}{0.0}{0}{19}

\draw[ultra thick,BrickRed] (7.05,0) -- (7.05,1.4);
\bound{7.05}{0.0}
\continuum{7.05}{0.0}{0}{14}

% YSR phase I
\draw[ultra thick,BrickRed] (7.90,0.4) -- (7.90,3.5);
\bound{7.90}{0.4}
\continuum{7.90}{0.4}{0}{31}

\draw[ultra thick,BrickRed] (8.75,0.0) -- (8.75,2.1);
\bound{8.75}{0.0}
\continuum{8.75}{0.0}{0}{21}

\draw[ultra thick,BrickRed] (9.55,0.0) -- (9.55,1.0);
\bound{9.55}{0.0}
\continuum{9.55}{0.0}{0}{9}

% YSR phase II
\draw[ultra thick,BrickRed] (10.5,0) -- (10.5,2.9);
\bound{10.5}{0.0}
\continuum{10.5}{0.0}{0}{29}

\draw[ultra thick,BrickRed] (11.25,0.4) -- (11.25,1.4);
\bound{11.25}{0.4}
\continuum{11.25}{0.4}{0}{10}

% bound-only tower
\draw[ultra thick,BrickRed] (12.0,0) -- (12.0,0.9);
\bound{12.0}{0.0}
\bound{12.0}{0.7}
\bound{12.0}{0.825}

% Local moment phase
\draw[ultra thick,BrickRed] (13.0,0) -- (13.0,2.6);
\bound{13.0}{0.0}
\continuum{13.0}{0.0}{0}{26}

\draw[ultra thick,BrickRed] (14.0,0) -- (14.0,1.7);
\bound{14.0}{0.0}
\continuum{14.0}{0.0}{0}{17}

\draw[ultra thick,BrickRed] (15.0,0) -- (15.0,0.7);
\bound{15.0}{0.0}
\bound{15.0}{0.7}

% ===== Top braces =====
\draw[decorate, decoration={brace, amplitude=10pt, raise=2pt}]
  (-0.0,4.65) -- node[above=14pt] {$\mathscr{PT}$-unbroken} (7.45,4.65);

\draw[decorate, decoration={brace, amplitude=10pt, raise=2pt}]
  (7.55,4.65) -- node[above=14pt] {$\mathscr{PT}$-broken} (12.45,4.65);

\draw[decorate, decoration={brace, amplitude=10pt, raise=2pt}]
  (12.55,4.65) -- node[above=14pt] {$\mathscr{PT}$-unbroken} (15.65,4.65);

% ===== Bottom braces =====
\draw[decorate, decoration={brace, amplitude=10pt, raise=2pt}]
  (9.85,-0.5) -- node[below=15pt] {Impurity is overscreened in the GS} (-0.0,-0.5);

\draw[decorate, decoration={brace, amplitude=10pt, raise=2pt}]
  (15.65,-0.5) -- node[below=15pt] {Impurity is unscreened in the GS} (10.25,-0.5);

\end{tikzpicture}
    \caption{
Phase diagram of the $n$-channel Kondo model with two spin-$\frac{1}{2}$ impurities coupled by a complex-conjugate interaction. The RG-invariant parameter $\alpha$ drives a sequence of Kondo, zero-mode, YSR, and local moment phases. The number and arrangement of excitation towers reflect the appearance of impurity string solutions and the transition from an overscreened to an unscreened impurity ground state at $\alpha=\frac{(n+1)\pi}{2}$. The Kondo, zero-mode, and local moment phases are $\mathscr{PT}$-unbroken, whereas the YSR phase is $\mathscr{PT}$-broken.
}
    \label{fig:phasediagram}
\end{figure*}
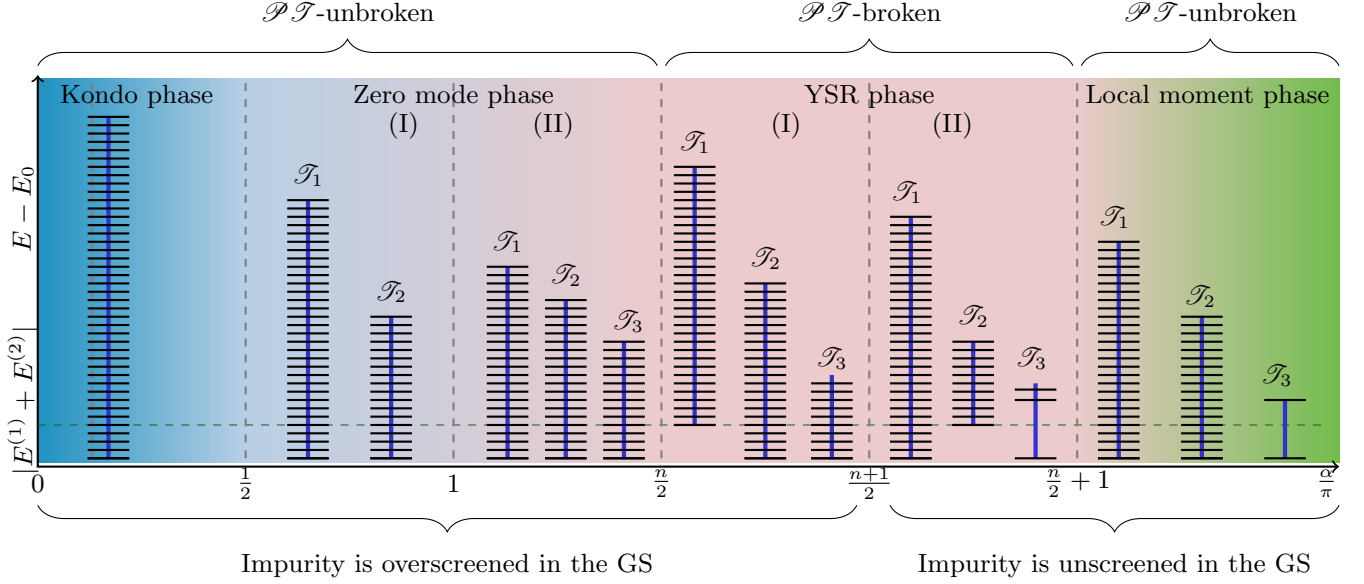

We now turn to discuss the phase diagram. When $0<\alpha<\frac{\pi}{2}$, corresponding to the {\em{overscreened Kondo phase}}, the solutions of the Bethe equations in the thermodynamic limit satisfy the string hypothesis and organize into $p$-strings,
\begin{equation}
\Lambda_{\gamma}^{(p,j)}
=
\Lambda_{\gamma}^{(p)}
+
\frac{ic}{2}(p+1-2j),
\qquad
j=1,\ldots,p,
\end{equation}
where $\Lambda_{\gamma}^{(p)}\in\mathbb{R}$ denotes the string center. In this phase, both impurities are overscreened by the multiparticle Kondo cloud formed by these bulk string excitations as shown in Fig.~\ref{fig:phasediagram}. The defect couplings are marginally relevant, generating the conventional monotonic RG flow from the ultraviolet to the infrared fixed point.

For $\frac{\pi}{2}<\alpha<\frac{n\pi}{2}$, corresponding to the {\em{zero\ mode phase}}, additional impurity-string solutions appear,
\begin{align}
\Lambda^{(1)}
&=
1-\cos\phi
+i\left(\sin\phi-\frac{c}{2}\right),\\
\Lambda^{(2)}
&=
1-\cos\phi
-i\left(\sin\phi+\frac{c}{2}\right),
\end{align}
whose energies vanish identically in the thermodynamic limit. In this phase, both impurities remain overscreened while supporting zero-energy impurity strings and the $\mathscr{PT}$ symmetry is unbroken. The zero-mode phase comprises two subphases. For $\frac{\pi}{2}<\alpha<\pi$ (zero mode I), only the fundamental impurity strings $\Lambda^{(1)}$ and $\Lambda^{(2)}$ are present. For $\pi<\alpha<\frac{n\pi}{2}$ ($\mathrm{zero\ mode\ II}$), higher-order impurity strings
\begin{equation}
\Lambda^{(\gamma,\ell)}
=
\Lambda^{(\gamma)}
-
ic\ell,
\qquad
\gamma=1,2,
\qquad
\ell=1,\ldots,p,
\end{equation}
also exist, where $p=\left\lfloor\frac{\alpha}{\pi}+\frac{1}{2}\right\rfloor$. In this phase, the RG flow connects the same ultraviolet and infrared fixed points, but the emergence of zero-energy impurity-string solutions reorganizes the excitation spectrum into multiple thermodynamic towers, rendering the $g$-function nonmonotonic.

For $\frac{n\pi}{2}<\alpha<\left(\frac{n}{2}+1\right)\pi$, corresponding to the {\em{YSR phases}}, the higher-order impurity strings persist, while the fundamental impurity strings acquire the complex energies
\begin{equation}
E^{(1)}
=
-2T_K e^{-i \left(\frac{\pi  n}{2}-\alpha \right)},
\quad
E^{(2)}
=
-2T_K e^{i \left(\frac{\pi  n}{2}-\alpha \right)},
\end{equation}
signaling spontaneous $\mathscr{PT}$-symmetry breaking. In the YSR-I phase ($n\pi/2<\alpha<(n+1)\pi/2$), the combined energy of the two fundamental impurity strings is negative, so the ground state contains them and the impurities are screened by these single-particle modes rather than by the multiparticle Kondo cloud. In the YSR-II phase ($ (n+1)\pi/2<\alpha<(\tfrac{n}{2}+1)\pi $), their combined energy is positive, leaving the impurities unscreened in the ground state but screened in the excited state containing the two impurity strings. Note that the zero-mode regime emerges only for multichannel systems ($n\ge2$); for $n=1$, the transition at $\alpha=\frac{\pi}{2}$ occurs directly between the Kondo and YSR phases as shown in Ref.~\cite{kattel2025dissipation,kattel2026monotonic}. The impurity strings acquire finite complex energies, introducing an intrinsic IR scale that destroys the conformal invariance.

Finally, for $\alpha>\left(\frac{n}{2}+1\right)\pi$ ({\em{local\ moment}} phase), the higher-order impurity strings continue to exist, and the energies of both the fundamental and higher-order impurity strings vanish, restoring
$\mathscr{PT}$ symmetry and leaving both impurities unscreened. The RG trajectory is cyclic, returning to the weak-coupling local-moment fixed point rather than connecting distinct ultraviolet and infrared fixed points. 

We study the finite-temperature thermodynamics in the $\mathscr{PT}$-unbroken phases, $0<\alpha<\frac{n\pi}{2}$ and $\alpha>\left(\frac{n}{2}+1\right)\pi$, where the impurities are respectively overscreened and unscreened. Introducing the particle and hole densities $\rho_p(\Lambda)$ and $\rho_p^h(\Lambda)$ of the $p$-string excitations and defining $\eta_p(\Lambda)=\frac{\rho_p^h(\Lambda)}{\rho_p(\Lambda)}$, the thermodynamic Bethe Ansatz equations take the usual recursive form
\begin{equation}
\ln\eta_p = - \frac{2D \delta_{p,n}}{T} \tan^{-1} \left(e^{\frac{\pi}{c}(\Lambda-1)}\right) + \sum_{\upsilon=\pm}G \ln(1+\eta_{p+\upsilon}),
\end{equation}
where
\begin{equation}
Gf(\lambda)=\int d\mu\,\frac{f(\mu)} {2\cosh\!\left[\pi(\lambda-\mu)\right]}.
\end{equation}
The hierarchy is supplemented by the boundary conditions $\eta_0(\lambda)=0$, and
\begin{equation}
\lim_{p\to\infty}\left\{[p+1]\ln\left(1+\eta_p\right)-[p]\ln\left(1+ \eta_{p+1}\right)\right\}=-\frac{h}{T},
\end{equation}
where $h$ denotes an external magnetic field. In what follows we set $h=0$.

Introducing the shifted rapidity
$\xi=\frac{\pi}{c}\Lambda-\ln\frac{T}{T_0}$ together with the scale
$T_0=De^{-\pi/c}$, yielding
\begin{equation}
\ln\eta_p
=
-2\delta_{p,n}e^\xi
+
G\left[
\ln(1+\eta_{p+1})
+
\ln(1+\eta_{p-1})
\right].
\end{equation}

The universal solution $\eta_p(\xi)$ interpolates between the ultraviolet ($\xi\to -\infty$) and infrared ($\xi\to \infty$) fixed points. Its asymptotic limits are
\begin{equation}
\eta_p(\xi\to -\infty)=(p+1)^2-1,\label{UV-limit}
\end{equation}
and
\begin{equation}
\eta_p(\xi\to\infty)=
\begin{cases}
\dfrac{\sin^2\left(\frac{(p+1)\pi}{n+2}\right)}
{\sin^2\left(\frac{\pi}{n+2}\right)}-1,& p<n,\\
(p+1-n)^2-1,& p\ge n.
\end{cases}
\label{IRlimit}
\end{equation}

The impurity contribution to the free energy depends on the parameter $\alpha$. When $0<\alpha<\frac{\pi}{2}$, there is a single excitation tower coming from each of the impurities. Following the usual TBA analysis~\cite{andrei1983solution}, two conjugate free energy contributions of both impurities sum up to

\begin{equation}
F_{\rm imp} = -\frac{T}{\pi} \int d\xi \frac{
\cos\alpha \cosh \left[\xi+\ln\left(\frac{T}{T_K}\right)\right]
\ln \left[1+\eta_1(\xi)\right]}{\cosh^2 \left[\xi+\ln\left(\frac{T}{T_K}\right)\right]-\sin^2\alpha}.\label{F imp Kondo}
\end{equation}

where 
\begin{equation}
T_K = T_0 \exp \left[ \frac{\pi}{c}(1-\cos\phi) \right] = D\exp \left[ -\frac{\pi}{c}\cos\phi \right].
\end{equation}

\begin{figure*}[t]
\centering

\includegraphics[width=0.48\textwidth]{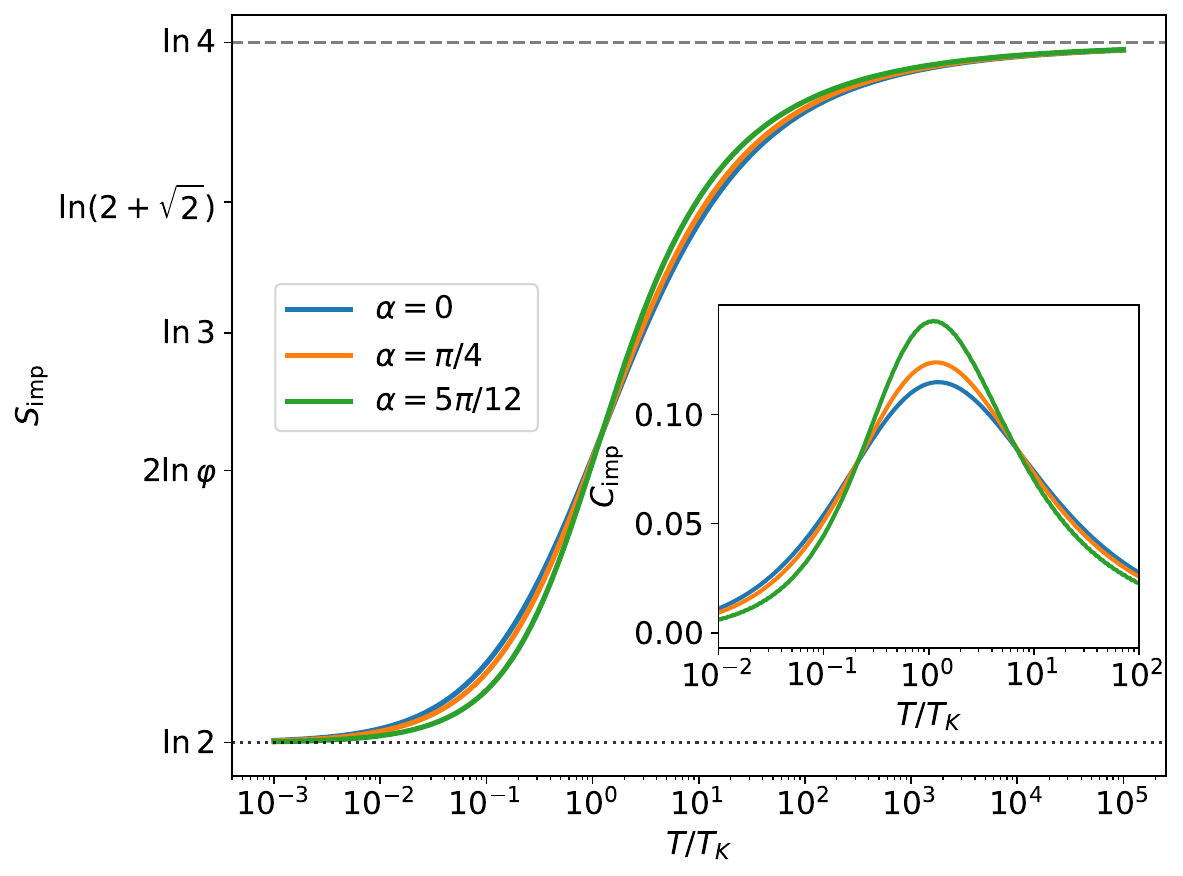}
\hfill
\includegraphics[width=0.48\textwidth]{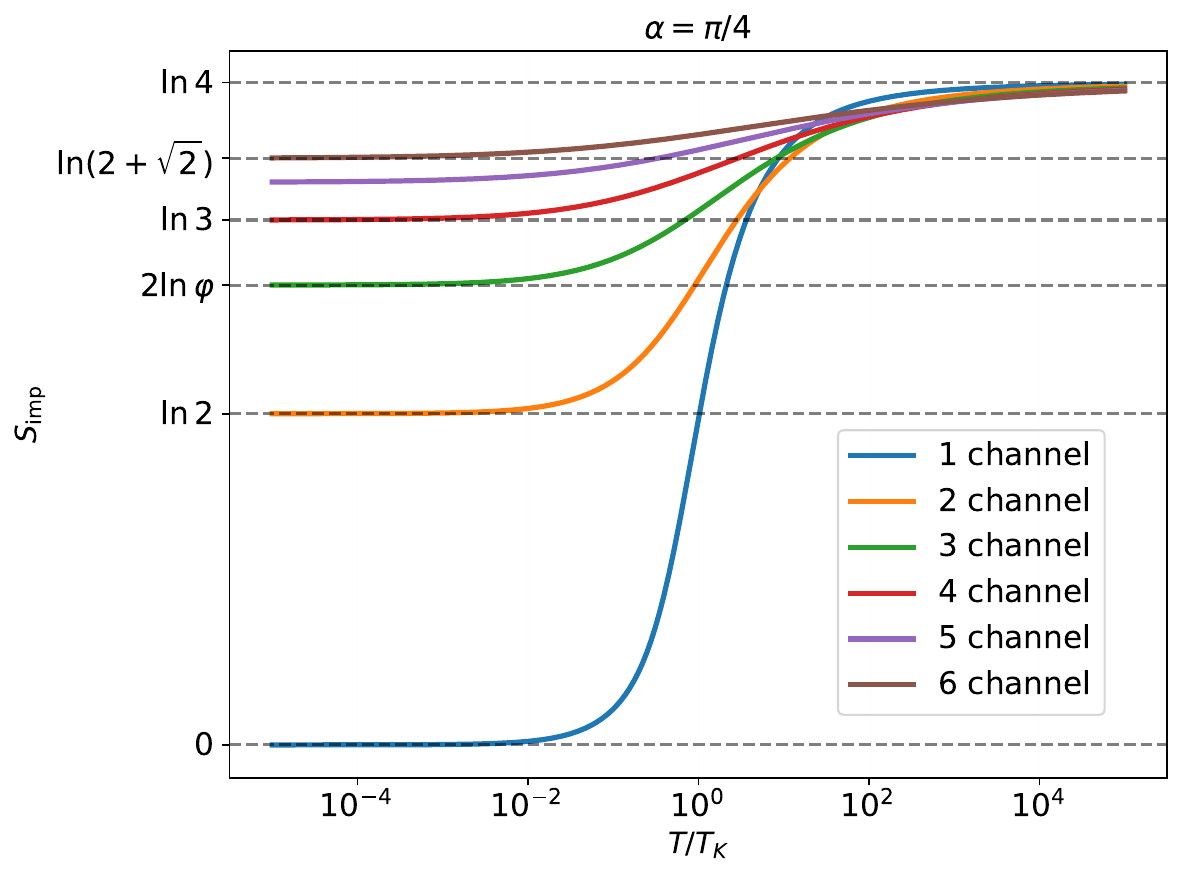}

\caption{Impurity entropy $S_{\rm imp}$ as a function of $T/T_K$. Left: two-channel model for several values of $\alpha$, with the corresponding impurity specific heat $C_{\rm imp}$ shown in the inset. Right: impurity entropy at fixed $\alpha=\pi/4$ for different channel numbers. Here $\varphi=(1+\sqrt{5})/2$ denotes the golden ratio.}
\label{fig:entropyheat}
\end{figure*}

Fig.~\ref{fig:entropyheat} shows the impurity entropy $S_{\rm imp}(T)=-\frac{\partial F_{\rm imp}(T)}{\partial T}
=\left(1+T\frac{\partial}{\partial T}\right)
\ln g_{\rm imp}(T)$ and impurity specific heat $C_{\rm imp}=T\,\partial S_{\rm imp}/\partial T$. Throughout the Kondo phase, $S_{\rm imp}$ decreases monotonically from the ultraviolet value $2\ln2$ to the infrared value $2\ln\left[2\cos\left(\frac{\pi}{n+2}\right)\right]$. The parameter $\alpha$ enters Eq.~\eqref{F imp Kondo} only through the width of a normalized kernel, reducing to the Hermitian kernel $1/\cosh$ at $\alpha = 0$ and narrowing to $\pi \delta[\xi + \ln(T/T_K)]$ as $\alpha \to \pi/2$. Increasing $\alpha$ therefore sharpens the crossover about $T_K$: raising the impurity entropy below $T_K$ and lowering it above, while the infrared entropy is determined solely by the number of channels.

For $\frac{\pi}{2}<\alpha<\pi$, the zero-energy impurity-string solutions reorganize the excitation spectrum into two towers. The first tower, $\mathscr T_1$, consists solely of the allowed bulk $p$-string excitations, whereas the second, $\mathscr T_2$, contains the impurity strings together with the allowed bulk strings.
The impurity partition function is obtained by summing the two tower contributions following the methods developed in Ref.~\cite{zhakenov2025thermodynamics,kattel2026thermodynamics,kattel2026multichannel}. The free energy contributions from the two towers of the first impurity are
\begin{align}
&F_{(2)}^{\mathscr{T}_1}(T) = -\frac{T}{2\pi} \int_{-\infty}^{\infty}
d\xi\, \frac{ \ln\left(1+\eta_1(\xi)\right)
}{\cosh\left(\xi+\ln\frac{T}{T_K}-i\alpha\right)},\\
&F_{(1)}^{\mathscr{T}_2}=F_{(2)}^{\mathscr{T}_1}-\frac{T}{2\pi}\int_{-\infty}^{\infty}
d\xi\,\frac{\ln\left(1+\eta_2(\xi)\right)}{\cosh\left(
\xi+\ln\frac{T}{T_K}-i\left(\alpha-\frac{\pi}{2}\right)\right)
}.
\end{align}

The corresponding second impurity free energies are related by
$\mathscr{PT}$ symmetry, $F_{(2)}^{\mathscr{T}_1}(T)=F_{(1)}^{\mathscr{T}_1}(T)^{*}$ and $F_{(2)}^{\mathscr{T}_2}(T)=F_{(1)}^{\mathscr{T}_2}(T)^{*}$. Each impurity therefore carries two towers, whose contributions are summed in its partition function, while the two impurities, being independent, contribute multiplicatively. Thus, the combined impurity partition function is
\begin{equation}
Z_{\rm imp} =\left( e^{-F_{(1)}^{\mathscr{T}_1}/T} + e^{-F_{(1)}^{\mathscr{T}_2}/T} \right)\left( e^{-F_{(2)}^{\mathscr{T}_1}/T} + e^{-F_{(2)}^{\mathscr{T}_2}/T} \right),
\end{equation}

such that the total combined free energy is
\begin{equation}
F_{\rm imp}(T) = -T\ln Z_{\rm imp}(T).
\end{equation}
Since the tower free energies of the two impurities are related by complex conjugation, $Z_{\rm imp, (2)} = Z_{\rm imp, (1)}^*$, so that $Z_{\rm imp} = |Z_{\rm imp, (1)}|^2 > 0$ and $F_{\rm imp}$ is real despite the individual tower contributions being complex. The same holds for the three-tower construction below.

\begin{figure*}
\centering

\includegraphics[width=0.48\textwidth]{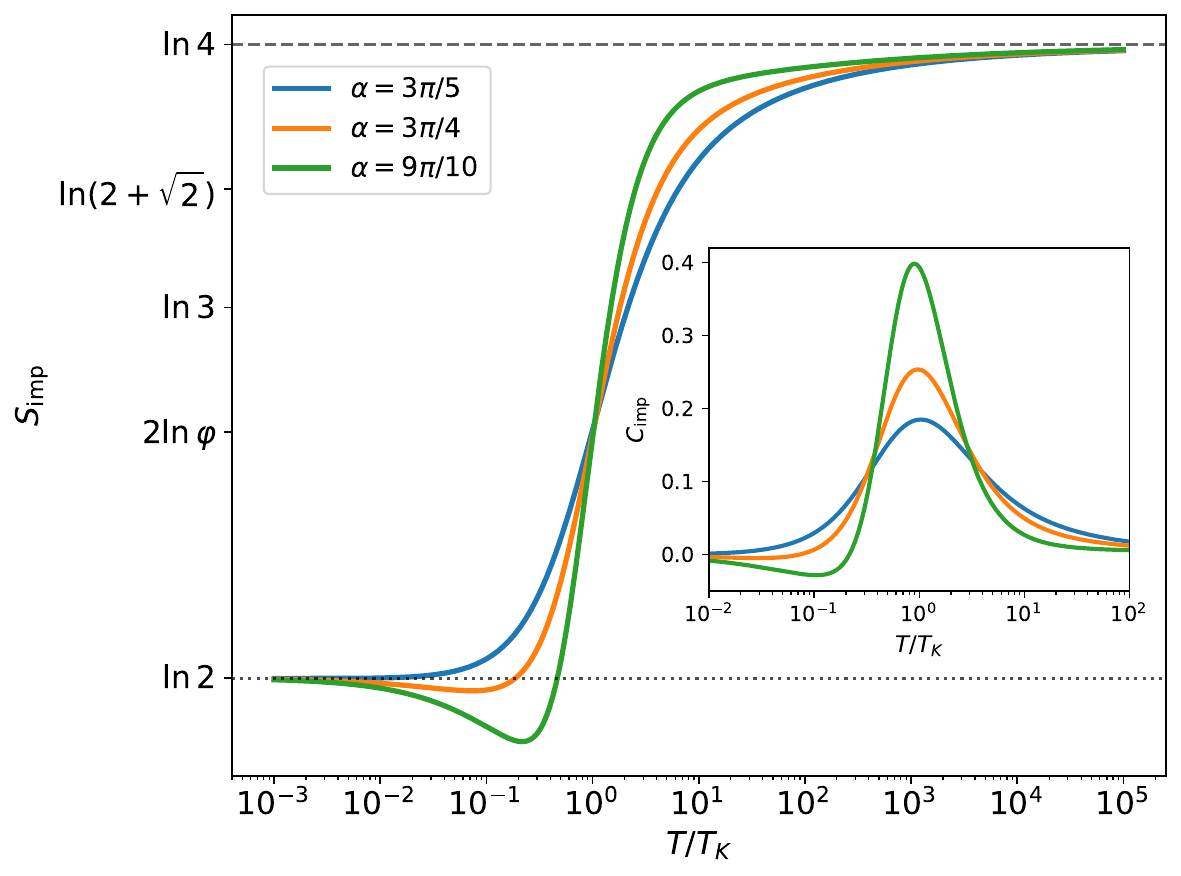}
\hfill
\includegraphics[width=0.48\textwidth]{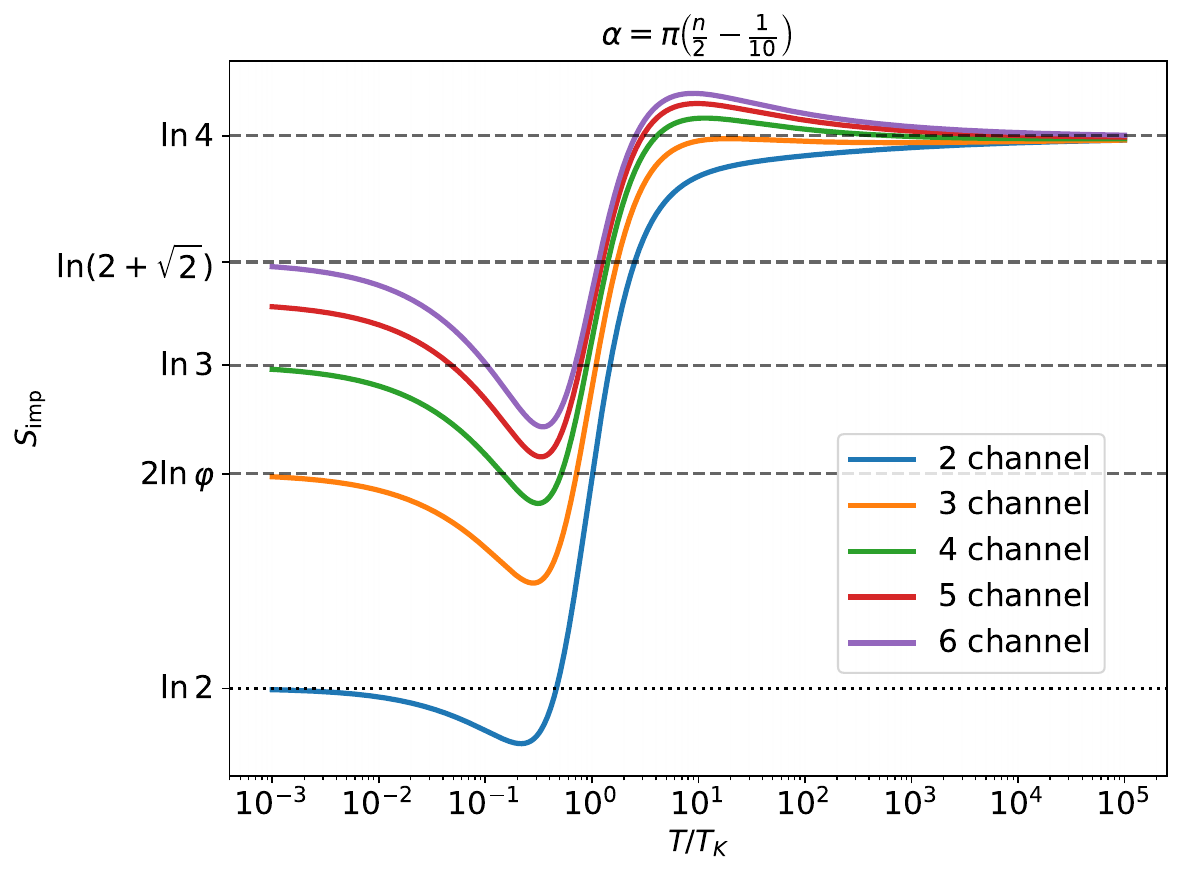}

\caption{
Impurity entropy $S_{\rm imp}$ as a function of $T/T_K$ in the zero-mode phase. Left: two-channel model for several values of $\alpha$, with the corresponding impurity specific heat $C_{\rm imp}$ shown in the inset. Increasing $\alpha$ continuously changes the entropy flow from monotonic to nonmonotonic. Right: impurity entropy for different channel numbers with $\alpha=\pi\left(\frac{n}{2}-\frac{1}{10}\right)$, for which the entropy is nonmonotonic for all channels shown.}
\label{fig:zero_mode_entropy}
\end{figure*}

For multichannel systems with $n\geq 3$, the zero mode II regime emerges for $\pi<\alpha<\frac{n\pi}{2}$, where higher-order zero-energy impurity strings reorganize the excitation spectrum into three towers. The first tower, $\mathscr T_1$, consists solely of the allowed bulk $p$-string excitations, the second, $\mathscr T_2$, contains the fundamental impurity strings together with the allowed bulk strings, while the third, $\mathscr T_3$, contains both the fundamental and higher-order impurity strings in addition to the allowed bulk strings. The impurity partition function is therefore obtained by summing over the contributions from all three towers. Defining $m=\lfloor\frac{2\alpha}{\pi}\rfloor$, the free energy contributions of the first impurity are
 \begin{subequations}\label{eq:tower-feng}
\begin{align}
F_{(1)}^{\mathscr{T}_1}(T) &= -\frac{T}{2\pi} \int_{-\infty}^{\infty}
d\xi \frac{\ln\left(1+\eta_{m+1}(\xi)\right)}{\cosh\left(\xi+\ln\frac{T}{T_K}-i\left(\alpha-\frac{m\pi}{2}\right)\right)}
\nonumber\\
&\quad
+\frac{T}{2\pi}\int_{-\infty}^{\infty}d\xi \frac{\ln\left(1+\eta_{m}(\xi)\right)}{\cosh\left(\xi+\ln\frac{T}{T_K}-i\left(\frac{(m+1)\pi}{2}-\alpha\right)\right)},\label{tower1-feng}\\
F_{(1)}^{\mathscr{T}_2}(T)&=\frac{T}{2\pi}\int_{-\infty}^{\infty}d\xi\frac{\ln\left(1+\eta_{m-1}(\xi)\right)}{\cosh\left(\xi+\ln\frac{T}{T_K}-i\left(\alpha-\frac{m\pi}{2}\right)\right)}
\nonumber\\
&\quad-\frac{T}{2\pi}\int_{-\infty}^{\infty}d\xi \frac{\ln\left(1+\eta_{m-2}(\xi)\right)}{\cosh\left(\xi+\ln\frac{T}{T_K}-i\left(\frac{(m+1)\pi}{2}-\alpha\right)\right)
},\label{tower2-feng}\\
F_{(1)}^{\mathscr{T}_3}(T)&=\frac{T}{2\pi}\int_{-\infty}^{\infty}d\xi\,
\frac{\ln\left(1+\eta_{m}(\xi)\right)}{\cosh\left(\xi+\ln\frac{T}{T_K}-i\left(\frac{(m+1)\pi}{2}-\alpha\right)\right)}\nonumber\\
&\quad+\frac{T}{2\pi}\int_{-\infty}^{\infty}d\xi\frac{\ln\left(1+\eta_{m-1}(\xi)\right)}{\cosh\left(\xi+\ln\frac{T}{T_K}-i\left(\alpha-\frac{m\pi}{2}\right)\right)}.\label{tower3-feng}
\end{align}
\end{subequations}

The second impurity free energies follow by $\mathscr{PT}$ conjugation, $F_{(2)}^{\mathscr{T}_\gamma}(T)=F_{(1)}^{\mathscr{T}_\gamma}(T)^{*},
\qquad
\gamma\in\{1,2,3\}$. The partition functions of the first and second impurities are
\begin{align}
Z_{(1)}(T)
=
\sum_{\gamma=1}^{3}
e^{-F_{(1)}^{\mathscr T_\gamma}(T)/T},\quad
Z_{(2)}(T)=
\sum_{\gamma=1}^{3}
e^{-F_{(2)}^{\mathscr T_\gamma}(T)/T},\nonumber
\end{align}
 hence, $Z_{\rm imp}(T)=Z_{(1)}(T)\,Z_{(2)}(T)$ and the impurity free energy is $F_{\rm imp}(T)=-T\ln Z_{\rm imp}(T)$.

Fig.~\ref{fig:zero_mode_entropy} shows the impurity entropy in the zero-mode phase. Although the ultraviolet and infrared fixed-point entropies remain unchanged, increasing $\alpha$ from $\pi/2$ toward $n\pi/2$ drives the impurity entropy from a monotonic to a nonmonotonic temperature dependence. The nonmonotonicity becomes more pronounced with increasing channel number. While the crossover from monotonic to nonmonotonic impurity entropy was previously observed numerically using the non-Hermitian numerical renormalization group~\cite{yi2026interplay,yi2026non}, the present exact Bethe Ansatz solution provides its analytical explanation by identifying the emergence of zero-energy impurity strings and the associated reorganization of the thermodynamic Bethe Ansatz into multiple excitation towers.

As discussed above, in the YSR phase the spontaneous breaking of $\mathscr{PT}$ symmetry renders the impurity spectrum complex, so that a straightforward thermodynamic Bethe Ansatz analysis is no longer applicable.

\begin{figure}
    \centering
    \includegraphics[width=1.0\linewidth]{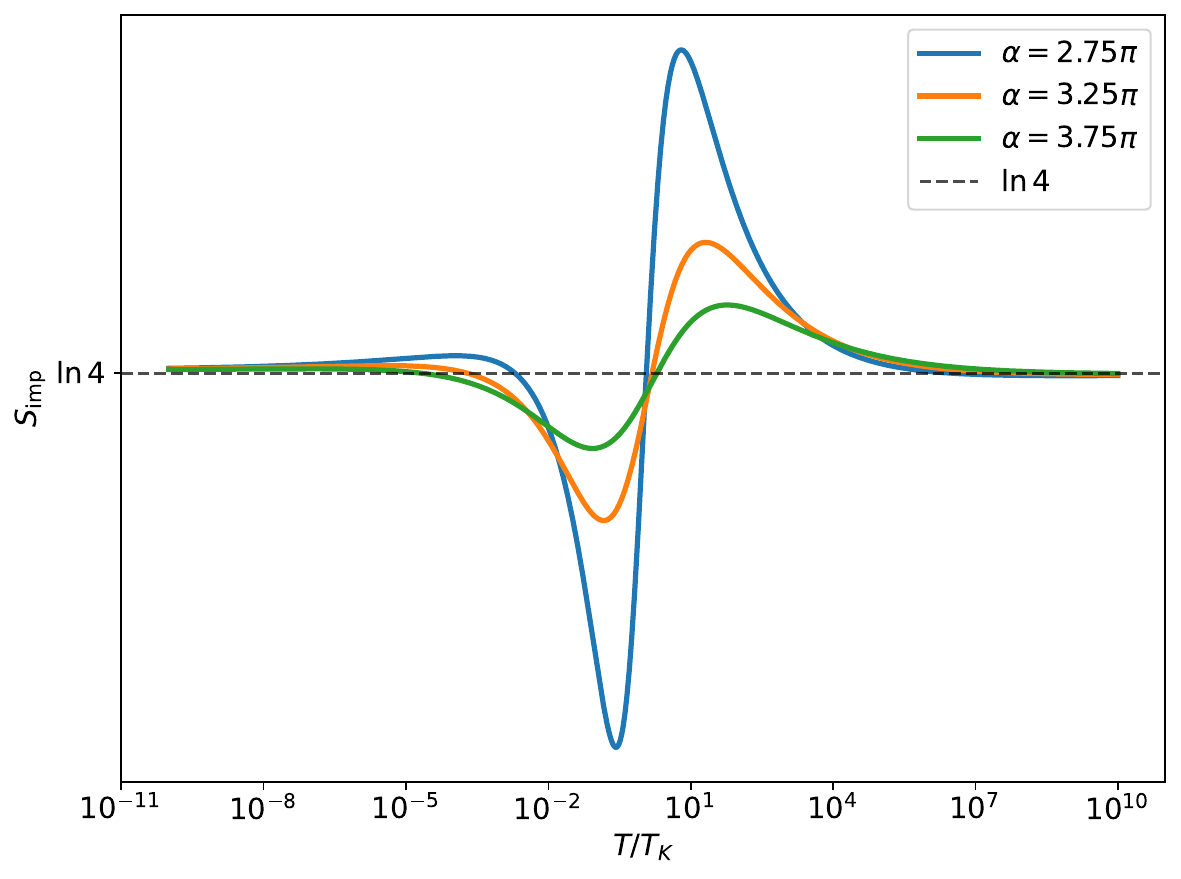}
    \caption{Impurity entropy $S_{\rm imp}$ as a function of $T/T_K$ in the local moment phase for the three-channel model. Both the UV and IR fixed points have $S_{\mathrm{imp}}=\ln4$, with $\alpha$-dependent overshooting and undershooting at intermediate temperatures.}
    \label{fig:local-moment}
\end{figure}

In the local-moment phase, $\mathscr{PT}$ symmetry is restored, and the energies of both the fundamental and higher-order impurity strings again vanish in the thermodynamic limit. Consequently, the impurity entropy is determined from the same excitation-tower construction as in the zero-mode phase, with the free-energy contributions given by Eq.~\eqref{eq:tower-feng}. Using the asymptotic solutions in Eqs.~\eqref{UV-limit} and \eqref{IRlimit}, the impurity entropy approaches $2\ln2$ in both the ultraviolet ($T\to\infty$) and infrared ($T\to0$) limits, consistent with the cyclic RG flow~\cite{nakagawa2018non,kattel2025dissipation,kattel2026monotonic}. As shown in Fig.~\ref{fig:local-moment}, the intermediate-temperature crossover exhibits both overshooting and undershooting before returning to $2\ln2$.

We have solved exactly a $\mathscr{PT}$-symmetric multichannel Kondo line defect by a generalized thermodynamic Bethe Ansatz. In the Kondo and zero-mode phases, the defect RG flow connects the same ultraviolet and infrared conformal defect fixed points as in the Hermitian theory, whereas the YSR phase spontaneously breaks $\mathscr{PT}$ symmetry and lies beyond the scope of our thermodynamic Bethe Ansatz. The local-moment phase instead exhibits cyclic RG flow, returning to the local-moment fixed point with impurity entropy $2\ln2$ in both the ultraviolet and infrared limits. The Kondo phase, where excitations are organized into a single thermodynamic tower, satisfies RG irreversibility, suggesting that a generalized Affleck--Ludwig $g$-theorem~\cite{affleck1991universal,friedan2004boundary,casini2016g} survives small departures from Hermiticity. By contrast, our exact solution of the zero-mode phase shows that a real spectrum and the correct ultraviolet and infrared defect CFT $g$-values are not sufficient: once zero-energy impurity strings reorganize the spectrum into multiple excitation towers, the $g$-function becomes nonmonotonic and the RG flow reversible. Establishing the precise conditions for such a generalized $g$-theorem remains an important open problem.

\textit{Acknowledgments:} We thank Colin Rylands for his careful review of the manuscript and for his valuable comments and insightful discussions. This work was supported by the Swiss National Science Foundation under Division II (Grant No.~200020-219400).

\bibliography{ref}

@article{kattel2025spin,
  title={Spin chain with non-Hermitian PT-symmetric boundary couplings: Exact solution, dissipative Kondo effect, and phase transitions on the edge},
  author={Kattel, Pradip and Pasnoori, Parameshwar R and Pixley, JH and Andrei, Natan},
  journal={Physical Review B},
  volume={111},
  number={22},
  pages={224407},
  year={2025},
  publisher={APS}
}

@article{sinha2026lattice,
  title={Lattice Topological Defects in Non-Unitary Conformal Field Theories},
  author={Sinha, Madhav and Tavares, Thiago Silva and Saleur, Hubert and Roy, Ananda},
  journal={arXiv preprint arXiv:2604.25999},
  year={2026}
}

@article{andrei1983solution,
  title={Solution of the Kondo problem},
  author={Andrei, Natan and Furuya, K and Lowenstein, JH},
  journal={Reviews of modern physics},
  volume={55},
  number={2},
  pages={331},
  year={1983},
  publisher={APS}
}

@article{tsvelick1983exact,
  title={Exact results in the theory of magnetic alloys},
  author={Tsvelick, AM and Wiegmann, PB},
  journal={Advances in Physics},
  volume={32},
  number={4},
  pages={453--713},
  year={1983},
  publisher={Taylor \& Francis}
}

@article{affleck1995conformal,
  title={Conformal field theory approach to the Kondo effect},
  author={Affleck, Ian},
  journal={arXiv preprint cond-mat/9512099},
  year={1995}
}

@article{affleck1991universal,
  title={Universal noninteger ‘‘ground-state degeneracy’’in critical quantum systems},
  author={Affleck, Ian and Ludwig, Andreas WW},
  journal={Physical Review Letters},
  volume={67},
  number={2},
  pages={161},
  year={1991},
  publisher={APS}
}

@article{friedan2004boundary,
  title={Boundary entropy of one-dimensional quantum systems at low temperature},
  author={Friedan, Daniel and Konechny, Anatoly},
  journal={Physical review letters},
  volume={93},
  number={3},
  pages={030402},
  year={2004},
  publisher={APS}
}

@article{nakagawa2018non,
  title={Non-Hermitian Kondo effect in ultracold alkaline-earth atoms},
  author={Nakagawa, Masaya and Kawakami, Norio and Ueda, Masahito},
  journal={arXiv preprint arXiv:1806.04039},
  year={2018}
}

@article{kattel2025dissipation,
  title={Dissipation driven phase transition in the non-Hermitian Kondo model},
  author={Kattel, Pradip and Zhakenov, Abay and Pasnoori, Parameshwar R and Azaria, Patrick and Andrei, Natan},
  journal={Physical Review B},
  volume={111},
  number={20},
  pages={L201106},
  year={2025},
  publisher={APS}
}

@article{yi2026non,
  title={Non-Hermiticity induced universal anomalies in Kondo conductance},
  author={Yi, Wei-Zhu and Chen, Yun and Pang, Jun-Jun and Chen, Hong and Wang, Baigeng and Wang, Rui},
  journal={Physical Review Letters},
  volume={136},
  number={11},
  pages={116502},
  year={2026},
  publisher={APS}
}

@article{yi2026interplay,
  title={Interplay between non-Fermi liquid and non-Hermiticity: A multimethod study of non-Hermitian multichannel Kondo model},
  author={Yi, Wei-Zhu and Chen, Yun and Pang, Jun-Jun and Chen, Hong and Wang, Baigeng and Wang, Rui},
  journal={Physical Review B},
  volume={113},
  number={16},
  pages={165110},
  year={2026},
  publisher={APS}
}

@article{burke2025non,
  title={Non-Hermitian numerical renormalization group: Solution of the non-Hermitian Kondo model},
  author={Burke, Phillip C and Mitchell, Andrew K},
  journal={Physical Review Letters},
  volume={135},
  number={20},
  pages={206502},
  year={2025},
  publisher={APS}
}

@article{zhakenov2025thermodynamics,
  title={Thermodynamics in a split Hilbert space: Quantum impurity at the edge of the Heisenberg chain},
  author={Zhakenov, Abay and Kattel, Pradip and Andrei, Natan},
  journal={arXiv preprint arXiv:2508.19334},
  year={2025}
}

@inproceedings{Liu2026ExtractingBC,
  title={Extracting Boundary Conformal Data from Periodic Non-Hermitian Critical Chains},
  author={Yifan Liu and Haruki Shimizu and Dongchang Liu and Kohei Kawabata},
  year={2026},
  url={https://arxiv.org/pdf/2606.16785}
}

@inproceedings{Maeda2026Holographic,
  title={Holographic Dual of PT Symmetric BCFT},
  author={Ryota Maeda and Nanami Nakamura and Tadashi Takayanagi },
  year={2026},
  url={https://arxiv.org/pdf/2606.18629}
}

@article{wilson1975renormalization,
  title={The renormalization group: Critical phenomena and the Kondo problem},
  author={Wilson, Kenneth G},
  journal={Reviews of modern physics},
  volume={47},
  number={4},
  pages={773},
  year={1975},
  publisher={APS}
}

@article{castro2017irreversibility,
  title={Irreversibility of the renormalization group flow in non-unitary quantum field theory},
  author={Castro-Alvaredo, Olalla A and Doyon, Benjamin and Ravanini, Francesco},
  journal={Journal of Physics A: Mathematical and Theoretical},
  volume={50},
  number={42},
  pages={424002},
  year={2017},
  publisher={IOP Publishing}
}

@article{casini2016g,
  title={The g-theorem and quantum information theory},
  author={Casini, Horacio and Landea, Ignacio Salazar and Torroba, Gonzalo},
  journal={Journal of High Energy Physics},
  volume={2016},
  number={10},
  pages={1--34},
  year={2016},
  publisher={Springer}
}

@article{gaiotto2021integrable,
  title={Integrable Kondo problems},
  author={Gaiotto, Davide and Lee, Ji Hoon and Wu, Jingxiang},
  journal={Journal of High Energy Physics},
  volume={2021},
  number={4},
  pages={268},
  year={2021},
  publisher={Springer}
}

@article{andrei1984solution,
  title={Solution of the multichannel Kondo problem},
  author={Andrei, N and Destri, C},
  journal={Physical review letters},
  volume={52},
  number={5},
  pages={364--367},
  year={1984},
  publisher={American Physical Society}
}

@article{tang2026exactly,
  title={Exactly solvable non-unitary conformal interfaces in unitary CFTs},
  author={Tang, Qicheng and Wei, Zixia and Wen, Xueda},
  journal={arXiv preprint arXiv:2606.32035},
  year={2026}
}

@article{furuta2026complex,
  title={Complex Conformal Manifolds},
  author={Furuta, Yuma and Harada, Wataru and Kusuki, Yuya and Tang, Yin},
  journal={arXiv preprint arXiv:2606.30720},
  year={2026}
}

@article{kattel2026thermodynamics,
  title={Thermodynamics in a split Hilbert space: Quantum impurity at the edge of a one-dimensional superconductor},
  author={Kattel, Pradip and Zhakenov, Abay and Andrei, Natan},
  journal={Physical Review B},
  volume={113},
  number={19},
  pages={195155},
  year={2026},
  publisher={APS}
}

@article{kattel2026multichannel,
  title={Multichannel Kondo effect in one-dimensional superconducting leads},
  author={Kattel, Pradip and Zhakenov, Abay and Andrei, Natan},
  journal={Physical Review B},
  volume={113},
  number={16},
  pages={165130},
  year={2026},
  publisher={APS}
}

@article{kattel2026monotonic,
  title={Monotonic Impurity Entropy beyond Unitarity: the $mathscr{PT}-$ Symmetric Quantum Impurity Model},
  author={Kattel, Pradip and Zhakenov, Abay and Andrei, Natan},
  journal={arXiv preprint arXiv:2606.28495},
  year={2026}
}

@article{yang1969thermodynamics,
  title={Thermodynamics of a one-dimensional system of bosons with repulsive delta-function interaction},
  author={Yang, Chen-Ning and Yang, Cheng P and others},
  journal={Journal of Mathematical Physics},
  volume={10},
  number={7},
  pages={1115},
  year={1969}
}

@article{takahashi1999thermodynamics,
  title={Thermodynamics of one-dimensional solvable models},
  author={Takahashi, Minoru},
  journal={ },
  volume={ },
  number={ },
  pages={ },
  year={1999},
}

@article{andrei1980diagonalization,
  title={Diagonalization of the kondo hamiltonian},
  author={Andrei, Natan},
  journal={Physical Review Letters},
  volume={45},
  number={5},
  pages={379},
  year={1980},
  publisher={APS}
}

@article{piquard2026experimental,
  title={Experimental Evidence of Fractional Entropy in Critical Kondo Systems},
  author={Piquard, C and Veillon, A and Sato, Y and Zanichelli, F and Aassime, A and Cavanna, A and Gennser, U and Mitchell, AK and Anthore, A and Pierre, F},
  journal={arXiv preprint arXiv:2605.00669},
  year={2026}
}

@article{child2022entropy,
  title={Entropy measurement of a strongly coupled quantum dot},
  author={Child, Timothy and Sheekey, Owen and L{\"u}scher, Silvia and Fallahi, Saeed and Gardner, Geoffrey C and Manfra, Michael and Mitchell, Andrew and Sela, Eran and Kleeorin, Yaakov and Meir, Yigal and others},
  journal={Physical Review Letters},
  volume={129},
  number={22},
  pages={227702},
  year={2022},
  publisher={APS}
}
\widetext
\section*{End Matter}

\subsection*{Derivation of Bethe Ansatz Equations}
Following the usual quantum inverse scattering, we compute the bare electron-impurity S-matrices. The electron scattering of one impurity gives
\begin{equation}
    S^{j0_1}=\frac{I-ice^{i\phi}P}{1-ice^{i\phi}}
\end{equation}
and likewise, the second impurity gives
\begin{equation}
    S^{j0_2}=\frac{I-ice^{-i\phi}P}{1-ice^{-i\phi}}
\end{equation}
The consistency condition requires the electron-electron S-matrix to be 
$S^{ij}=P$ and the transfer matrix with periodic boundary conditions becomes
\begin{align}
\left(S^{j j-1} \ldots S^{J0_1}\ldots S^{j 1} S^{j N} \ldots S^{J0_2} \ldots S^{j j+1}\right)_{a_{1} \ldots a_{N}}^{b_{1} \ldots b_{N}}A_{b_1\cdots b_N} e^{i k_j L}= A_{a_1\cdots a_N}
\label{pbcc}
\end{align}
We define a continuous version of the scattering matrix
\begin{equation}
S(u)=\frac{u I+i c P}{u+i c} \equiv a(u) I+b(u) P
\end{equation} 
where the parameter $u$ is called the spectral parameter. When $u=u_j$, this reduces to the correct S-matrix if we identify $u_j=1$ for electrons and $u_j=0$ for the first impurity and $u_j=1-e^{2i\phi}$ for the second impurity. This S-matrix satisfies a continuous version of the Yang-Baxter relation
\begin{equation}
S^{k j}(u-v) S^{k i}(u) S^{j i}(v)=S^{j i}(v) S^{k i}(u) S^{k j}(u-v)
\label{contybe}
\end{equation}
We introduce a fictitious particle in an auxiliary space and scatter it through all the particles. Finally, we define the monodromy matrix as the product of all such scattering
\begin{equation}
\Xi(u)=\prod_{y=1}^N S^{ya}(u-u_y)\, ,
\end{equation}
where the index $a$ denotes an auxiliary particle.
Taking the trace over the auxiliary space, we define the transfer matrix as
\begin{equation}
\mathcal{T}(u)\equiv Tr_a{\Xi}(u)=S^{j j-1}(u-u_{j-1}) \cdots S^{j 2}(u-u_2) S^{j 1}(u-u_1) S^{j N}(u-u_N) \cdots S^{j j+1}(u-u_{j+1})
\label{transfermat}
\end{equation}
Introducing $R$ matrix defined as
\begin{equation}
R=S(u-v) P=\frac{(u-v) P+i c I}{(u-v)+i c}
\end{equation} 
Eq.\eqref{contybe} implies 
\begin{equation}
R_{p, w}^{s, t} S(u)_{a, s}^{d, q} S(v)_{d, t}^{b, z}=S(v)_{a, p}^{c, s^{\prime}} S(u)_{c, w}^{b, t^{\prime}}R_{s^{\prime}, t^{\prime}}^{q, z}
\label{RSSeqn}
\end{equation}
Since the monodromy matrix is the product of the scattering matrix, by the repeated application of Eq.\eqref{RSSeqn}, we conclude
\begin{equation}
R_{p, w}^{s, t} \Xi(u)_{s}^{q} \Xi(v)_{t}^{z}=\Xi(v)_{p}^{s^{\prime}} \Xi(u)_{w}^{t^{\prime}} R_{s^{\prime}, t^{\prime}}^{q, z}
\end{equation}
Writing the above equation as
\begin{equation}
\Xi(u)_{s}^{q} \Xi(v)_{t}^{z}=\left(R_{p, w}^{s, t}\right)^{-1} \Xi(v)_{p}^{s^{\prime}} \Xi(u)_{w}^{t^{\prime}} R_{s^{\prime}, t^{\prime}}^{q, z}
\label{rsisi}
\end{equation}
Taking the trace over the auxiliary space, we get
\begin{equation}
[\mathcal{T}(u),\mathcal{T}(v)]=0
\end{equation}
The existence of this infinite number of conserved quantities shows that the model is integrable.

Using the usual quantum inverse scattering approach, the Bethe Ansatz equations of the model with one channel of conduction electron can be obtained as
\begin{align}
e^{i k_{j} L}&=\prod_{\gamma=1}^{M} \frac{\Lambda_{\gamma}-1+i ce^{i\phi} / 2}{\Lambda_{\gamma}-1-i ce^{i\phi} / 2}\label{momentacond}\\
-\prod_{\delta=1}^{M} \frac{\Lambda_{\delta}-\Lambda_{\gamma}+i ce^{i\phi}}{\Lambda_{\delta}-\Lambda_{\gamma}-i ce^{i\phi}}&=\left(\frac{\Lambda_{\gamma}-1-i ce^{i\phi} / 2}{\Lambda_{\gamma}-1+i ce^{i\phi} / 2}\right)^{N^{e}}\left(\frac{\Lambda_{\gamma}-i ce^{i\phi}/ 2}{\Lambda_{\gamma}+i ce^{i\phi}/ 2}\right)\left(\frac{\Lambda_{\gamma}-(1-e^{2i\phi})-i ce^{i\phi}/ 2}{\Lambda_{\gamma}-(1-e^{2i\phi})+i ce^{i\phi}/ 2}\right)\label{unwantedterms}
\end{align}

These Bethe Ansatz equations can be made more symmetric upon changing the variable $\Lambda_\gamma\to e^{i\phi}(\Lambda_\gamma-1)+1$ such that the equations become

\begin{align}
e^{i k_{j} L}&=\prod_{\gamma=1}^{M} \frac{\Lambda_{\gamma}-1+i\frac{c}{2}}{\Lambda_{\gamma}-1-i \frac{c}{2}}\label{CBAE}\\
\prod_{\delta=1, \delta\neq \gamma}^{M} \frac{\Lambda_{\delta}-\Lambda_{\gamma}+i c}{\Lambda_{\delta}-\Lambda_{\gamma}-i c}&=\left(\frac{\Lambda_{\gamma}-1-i \frac{c}{2}}{\Lambda_{\gamma}-1+i \frac{c}{2}}\right)^{N^{e}}\left(
\frac{\Lambda_{\gamma}-1+e^{-i\phi}-i\frac{c}{2}}
     {\Lambda_{\gamma}-1+e^{-i\phi}+i\frac{c}{2}}
\right)
\left(
\frac{\Lambda_{\gamma}-1+e^{i\phi}-i\frac{c}{2}}
     {\Lambda_{\gamma}-1+e^{i\phi}+i\frac{c}{2}}
\right)\label{SBAE}
\end{align}
In the equation above we excluded the term with $\delta=\gamma$ from the product on the left-hand-side to cancel out the minus sign in the Eq.~\eqref{unwantedterms}.

To obtain the equations for the model with a generic $n-$flavor of conduction electron, we invoke the dynamical fusion system, which was systematically developed in Ref.~\cite{andrei1984solution}, we obtain the Bethe Ansatz equation as
\begin{equation}
e^{i k_{j} L} =\prod_{\gamma=1}^{M} \frac{\Lambda_{\gamma}-1+i \frac{cn}{2}}
     {\Lambda_{\gamma}-1-i \frac{cn}{2}},\label{CBAE-multch}
\end{equation}

and 

\begin{equation}
    \prod_{\delta=1,\delta\neq\gamma}^{M}
\frac{\Lambda_{\delta}-\Lambda_{\gamma}+ic}
     {\Lambda_{\delta}-\Lambda_{\gamma}-ic}
=
\left(
\frac{\Lambda_{\gamma}-1-i\frac{cn}{2}}
     {\Lambda_{\gamma}-1+i\frac{cn}{2}}
\right)^{N^e}\left(
\frac{\Lambda_{\gamma}-1+e^{-i\phi}-i\frac{c}{2}}
     {\Lambda_{\gamma}-1+e^{-i\phi}+i\frac{c}{2}}
\right)
\left(
\frac{\Lambda_{\gamma}-1+e^{i\phi}-i\frac{c}{2}}
     {\Lambda_{\gamma}-1+e^{i\phi}+i\frac{c}{2}}
\right),\label{SBAE-multch}
\end{equation}

\end{document}